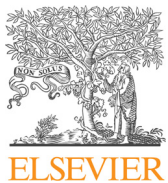
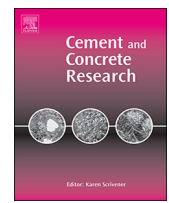

# Producing C-S-H gel by reaction between silica oligomers and portlandite: A promising approach to repair cementitious materials

Rafael Zarzuela[a], Manuel Luna[a], Luis M. Carrascosa[a], María P. Yeste[b,c], Inés Garcia-Lodeiro[d], M. Teresa Blanco-Varela[d], Miguel A. Cauqui[b,c], José M. Rodríguez-Izquierdo[b,c], María J. Mosquera[a,*]

[a] *TEP-243 Nanomaterials Group, Department of Physical-Chemistry, Faculty of Sciences, University of Cadiz, 11510 Puerto Real, Spain*
[b] *Department of Materials Science, Metallurgical Engineering & Inorganic Chemistry, Faculty of Sciences, University of Cadiz, Campus Río San Pedro s/n, 11510 Puerto Real, Cádiz, Spain*
[c] *IMEYMAT, Institute of Electron Microscopy and Materials, University of Cadiz, Spain*
[d] *Eduardo Torroja Institute for Construction Science (IETcc-CSIC), c/ Serrano Galvache 4, 28033 Madrid, Spain*

ABSTRACT

Impregnation treatments are one of the alternatives to protect concrete-based building and monuments from weathering degradation. However, it is important to consider the chemical compatibility of the reaction products with the building material. The impregnation product studied here consists of a silica oligomer able to polymerize, by a simple sol-gel process, inside the pore structure of concrete. In this work, we investigate the ability of this impregnation treatment to produce C-S-H gel in contact with cement paste. A complete characterization of the reaction products demonstrated that the silanol groups from silica oligomers react with the portlandite present in the cement paste generating a material with the chemical, structural and morphological features of C-S-H gel. Simultaneously, the $^{29}$Si NMR results indicate that the Si—O units are incorporated into the existing C-S-H, increasing its chain length. These results open the way for a simple concrete structures repairing procedure.

## 1. Introduction

The technological advances brought by the procedure to obtain Portland cement and the use of reinforced concrete shaped late 19th and 20th century architecture, giving rise to the architectural styles of the modern era (modernism, rationalism, brutalism…). As a defining characteristic of said styles, these new reinforced structures were often left in fair-faced condition [1], without coatings or painting, increasing its susceptibility to suffer degradation under aggressive environments. The earlier manifestation of most degradation processes is the onset of nano-and micro-cracks that is usually followed by increased porosity [2], material loss and decreased mechanical performance [3]. In addition to these phenomena, reinforced concrete is specially affected by the ingress of soluble $CO_2$ and $Cl^-$ ions, which may cause de-passivation and corrosion of the reinforcements, eventually resulting in delamination and structural failure. Nowadays, the current manufacturing processes and construction techniques are designed to minimize these issues. However, in the case of older concrete-based structures the limited knowledge about the concrete degradation mechanisms has given rise to a sizeable number of buildings that require repair interventions and preventive maintenance treatments to combat their damages.

Nowadays, the application of surface treatments, which prevent the water ingress and fill the cracks produced by degradation, is the preferred choice for preserving historic concrete [4]. Although coatings such as epoxy and acrylic resins are commonly employed with this aim, they present significant drawbacks associated to thick layer formation (0.1–1 mm) on concrete [5]. These layers can modify the aesthetic conditions of the surface and, especially, they present low durability due to low penetration and poor adhesion to the substrate. Impregnation treatments, on the other hand, present a low viscosity, and consequently have the ability to penetrate deeper (1–20 mm) and react in situ within the pore structure of the damaged substrate, promoting long-lasting effectiveness [5]. Different treatments based on nano-silica dispersions, available in the market, are employed for concrete protection due to their ability to undergo pozzolanic reaction with the portlandite and fill the pores with C-S-H. However, despite the small particle size, these treatments have difficulties penetrating into the smaller pores [5,6].

Silica oligomer/monomer-based sols (e.g., tetraethoxysilane, TEOS)

are widely used in the stone conservation market [7] due to their advantages and their economical cost comparable to other alternatives. Their low viscosity allows them to penetrate deeply into the porous structure and after polymerization, which occurs upon reaction with environmental moisture, a stable gel with a silicon-oxygen backbone is formed. This $SiO_2$ structure has a similar composition to the quartz and silicate minerals present in a wide variety of building materials, including the hydrated calcium silicate (C-S-H) and aluminosilicates (e.g. strätlingite) present in concrete or the siliceous sands used as fine aggregate. As the main drawbacks for these products, xerogel cracking is produced due to the high capillary pressures inside the dense microporous gel network and their effectiveness tends to be limited on carbonate-rich materials, such as limestone and concrete surfaces subjected to intense carbonation.

In previous developments [8,9], we devised an ultrasound-assisted synthesis to produce crack-free mesoporous xerogels by adding a small proportion (< 0.5%) of a long chain amine which acts as a catalyst and surfactant [10,11]. This synthesis process has been patented [12] and is currently under exploitation. The ultrasound agitation produces a homogeneous emulsion of water in ethoxysilanes in a short time (< 10 min), which is stabilized by the surfactant via formation of inverse micelles. The micelles created by the surfactant act as nanoreactors, where the ethoxysilane hydrolyzes and polymerizes forming $SiO_2$ nuclei that eventually grow outside the micelles forming a gel composed of densely packed nanoparticles with uniform size. The mesoporous structure of the gel (6–20 nm pores) decreases capillary pressure during drying and prevents cracking. An additional advantage of this synthesis route regards the higher compatibility of the treatment with carbonate-based materials, as observed in previous works [13].

The application of silica oligomer/monomer-based products as an impregnation treatment on concrete has been hardly investigated, despite recent studies indicating their higher effectiveness in comparison to other impregnation products (such as sodium silicate and nanosilica) [14,15]. However, the high pH and different composition of the cement matrix compared to stone is likely to modify the interaction of these products with concrete and thus, it requires to be investigated.

The effectiveness demonstrated by silica-based sols to repair degraded concrete structures can be clearly associated to their reactivity in contact with cementitious substrates. It is well known that calcium-silicate-hydrate (C-S-H) gel is a significant component in the cement paste, giving rise to its engineering properties, and durability [16]. New C-S-H gel could be produced into the concrete porous system as consequence of reaction between silanol groups from silica-based products and $Ca^{2+}$ ions from portlandite, one of the main resultant products of cement hydration. Furthermore, since the silica source is already solubilized, this process can be faster than the pozzolanic reaction with other (solid) silica sources such as nano-$SiO_2$. In spite of the significance of this feature, that would allow to repair the concrete buildings and structures by a simple and low cost impregnation treatment, scarce research about that can be found in the literature [17,18].

Starting from this premise, the objective of this work is to investigate the ability of the silica oligomer-based impregnation, product previously developed and patented by our group [10,12], to produce C-S-H gel. For this purpose, the sol-gel reaction was carried out in the presence of portlandite paste. This system allows to simulate the conditions of the concrete pore solution for studying the formation of C-S-H gel, eliminating the interferences introduced by the C-S-H and other silica phases of hydrated cement present originally in the cement paste. An extensive structural, chemical and morphological characterization of the reaction products was carried out. Once the C-S-H formation from portlandite was evidenced, the study was carried out by using cement paste in order to be brought closer to real conditions.

## 2. Materials and methods

### 2.1. Synthesis of the impregnation product

The product employed for this study (henceforth called UCA-T) was synthetized according to a previously developed surfactant-assisted sol–gel method developed and patented by our group [10,12]. The starting sol was prepared by mixing 100 ml of an oligomeric silica precursor (TES40 WN, Wacker Chemie AG) with 0.16 ml n-octylamine (99%, Sigma-Aldrich), which acts as a catalyst and a surfactant, and 0.5 ml de-ionized $H_2O$. According to its technical data sheet, TES40 is a mixture of monomeric and oligomeric ethoxysilanes with linear chains of 4–5 Si−O units and a 41% $SiO_2$ content. The mixture was afterwards subjected to ultrasonic agitation for 10 min utilizing an ultrasound probe (Bandelin Ultrasonic HD3200) working at 74% amplitude in a water bath to prevent overheating.

Xerogel samples were obtained, in order to study the product properties, in absence of cement paste, by casting 15 ml the sol in an open Ø85 mm Petri dish and keeping them under laboratory conditions (20 °C, 40% RH) for gelification. Once the samples reached constant weight, the xerogel was ground for characterization.

### 2.2. Reaction of the product with portlandite

A first approach to verify the ability of the product to form C-S-H gel, was to study its reaction with portlandite paste (PP). The samples for this study were prepared as follows: (1) A PP was obtained by mixing, under mechanical stirring, $Ca(OH)_2$ (90%, Scharlau) and de-carbonated $H_2O$ in a water/solid ratio of 0.5 (2) 2 g of the UCA-T sol were mixed with 15 g of the PP (3) The pastes were placed in desiccator under a $N_2$ atmosphere with saturated humidity and left to cure (4) At different time intervals (7 and 21 days), the reaction was stopped by replacing the water via solvent exchange with isopropanol [19]. (4) Isopropanol was evaporated in a vacuum oven at 35 °C during three days. For comparison purposes, a portlandite paste without UCA-T was also subjected to the same process (in this case, the paste was cured for 7 days).

### 2.3. Reaction of the product with hydrated cement paste

The second series of experiments focused on studying the interaction of the impregnation product with the hydration products present in concrete. To avoid the interferences due to the aggregates, the experiments were carried out using hydrated cement paste (CP), prepared by mixing a Cem I 42.5 R cement with de-carbonated $H_2O$ in a 0.5 w/c ratio. The paste was cured in a $N_2$ atmosphere saturated with water for 28 days, and afterwards vacuum-dried at 35 °C during three days.

Prior to mixing with UCA-T, the cement paste was finely ground to a size of < 30 μm. 15 g of the powdered cement paste was mixed with 3 g of the UCA-T sol and stored in a $N_2$ atmosphere saturated with ambient water. After curing for 7 and 21 days, the hydration reactions were stopped by replacing the water via solvent exchange with isopropanol and, afterwards, dried in a vacuum oven at 35 °C during three days. For comparison purposes, a sample of CP was also subjected to the same experimental conditions.

### 2.4. Characterization of the samples

Structural, chemical and morphological characterization of the xerogels and the reaction products of the sol with portlandite and cement paste was carried out by the techniques described below. Prior to their analysis, all the solids under study were finely ground (< 30 μm).

Mineralogical and chemical analysis of the samples was performed by X-ray diffraction (XRD), X-ray fluorescence (XRF) and thermogravimetric analysis (TGA) techniques. The XRD patterns of the materials were obtained with a Bruker D8 advance diffractometer equipped with a secondary monochromator, Cu tube X-ray, using Cu Kα radiation. Quantitative analysis of the phases was carried out by the Rietveld method. Compositional analysis was carried out by micro-XRF (μ-XRF) analysis, using a Bruker S4-PIONEER equipment. The sample compositions were obtained averaging the XRF elemental maps of the powdered samples spread on a plate. TGA was carried out in a SDT Q600 TA-Instruments thermobalance. The samples were heated up to 800 °C at 5 °C/min, in a flow of $O_2(5\%)/He$ (60 cm$^3$/min). Such TGA experiments were complemented by analyzing the released gases using a quadrupole mass spectrometer (Pfeiffer, QME-200-D-35614 model) to track MS signals and identify the desorption and decomposition products.

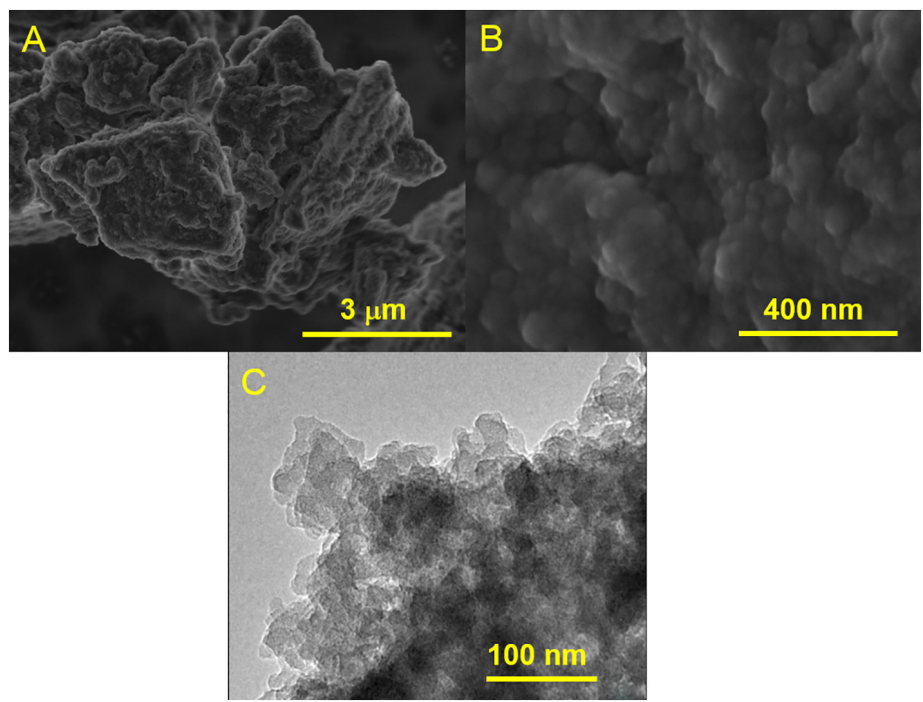

**Fig. 1.** (A, B) SEM micrographs of the UCA-T xerogel. (C) TEM micrograph of the UCA-T xerogel.

Structural characterization of the materials was performed by Fourier-transform infrared spectroscopy (FTIR) and $^{29}$Si nuclear magnetic resonance ($^{29}$Si NMR). The samples for the FTIR measurements were prepared in KBr pellets (1% proportion) and measured in transmittance mode in the 4000–450 cm$^{-1}$ range by using a IRAffinity-1S spectrophotometer from Shimadzu. In order to perform semi-quantitative analysis of the portlandite and carbonates, $K_3Fe(CN)_6$ was added ($K_3Fe(CN)_6$/sample ratio 1:1) as an internal standard. Single-pulse (SP) MAS-$^{29}$Si NMR experiments were carried out on a Bruker ADVANCE WB400 spectrometer equipped with a multinuclear probe. Samples of powdered materials were packed in 4 mm zirconia rotors and spun at 8 kHz. Pulse length was 5 μs, with a recycle delay of 30 s and the number of transients was 8000. The chemical shift values are expressed in ppm of tetramethylsilane.

Morphological features of the materials were studied by electron microscopy. Field Emission Scanning Electron Microscopy (FE-SEM) micrographs were registered in secondary electrons mode with a NanoSEM 450 model from the FEI Company, working at an acceleration voltage of 2.5 kV. The samples were prepared by directly depositing a small amount of the powder over carbon tape placed in an aluminum sample holder. Transmission Electron Microscopy (TEM) and Scanning Transmission Electron Microscopy Bright Field (STEM-BF) images, were obtained with a FEI TITAN3 Themis 60–300 microscope, operating at an acceleration voltage of 60 kV and equipped with an energy-dispersive X-ray spectroscopy EDXS detector for selected area EDXS analysis. The samples were prepared by finely grounding the powders and direct deposition over a lacey carbon-coated copper grid.

### 3. Results and discussion

#### 3.1. Characterization of the impregnation product

Characterization of the UCA-T sol and the sol-gel transition are reported elsewhere [10]. A summary of these properties (density, viscosity, surface tension, gel time and stability) can be found in supporting information (Table S1). When stored in a closed vessel, the sol remains stable (i.e. does not gel or change its properties), as the amount of water is not enough to self-sustain the hydrolysis of the silica precursor. Upon exposure to ambient moisture (20 °C, 40%RH), gel time is between 16 and 17 h, which is long enough to allow manipulation of the product in realistic application conditions.

After the sol-gel transition, UCA-T forms a monolithic xerogel due to the effect of the surfactant (n-octylamine), which plays two important roles: (i) it reduces the surface tension and (ii) it promotes the formation of a particulated xerogel with a mesoporous structure. This combination of surface tension reduction and increasing of pore size results in a lowering of capillary pressure that allows monolithic xerogels to be obtained [10].

The final weight (after 60 days) of the xerogel was a 65% of the initial (sol) weight, a value remarkably higher than the theoretical weight that would result from a fully realized sol-gel reaction to produce $SiO_2$ (~40%), indicating the presence of unreacted ethoxy- groups and/or byproducts ($H_2O$ and EtOH) trapped inside the smaller pores of the structure.

The SEM and TEM images reveal the mesoporous structure of the UCA-T xerogel. Specifically, the SEM micrographs (Fig. 1A, 1B) show a globular structure, composed of a network of tightly packed $SiO_2$ nanoparticles with a uniform size, in line with previous observations where it was found that the surfactant produced inverse micelles that act as nano-reactors promoting the formation of a mesostructure [10]. The detailed micrographs of the particle structure, obtained by TEM (Fig. 1C), show a uniform size distribution of the nanoparticles in the 20–50 nm range, with the minor presence of some particles around 70 nm. The uniform particle size clearly confirms the role played by the inverse micelles as nano-reactors of the mesostructured. The particle morphology was an irregular quasi-spherical shape.

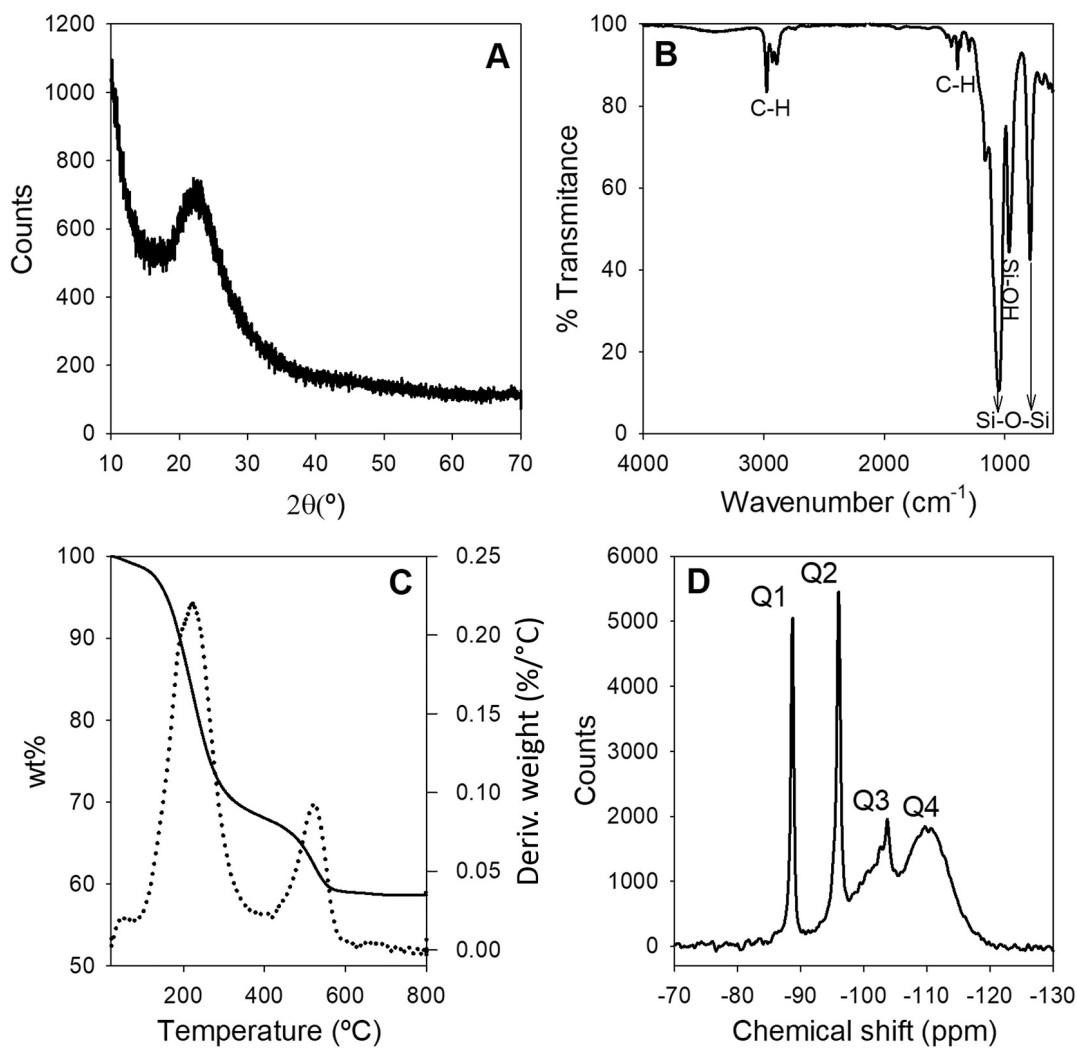

**Fig. 2.** Characterization of the UCA-T xerogel. (A) XRD pattern. (B) Attenuated transmission reflectance (ATR)-FTIR spectrum. (C) TGA curve. (D) $^{29}$Si NMR spectra.

The amorphous nature of the UCA-T was confirmed by the XRD pattern (Fig. 2A), which shows the characteristic wide and poorly defined band of non-crystalline structures. The position of the band, centered around $2\theta = 22°$, is typically found in the amorphous silica gels and corresponds to the distance between Si atoms bridged by O atoms [20,21]. The composition of the xerogel, according to the XRF analysis, was predominantly $SiO_2$ (99.5%) with a minor contribution of Ca and Al impurities.

The chemical nature of the UCA-T xerogel was further evaluated by infrared spectroscopy. The FTIR spectra (Fig. 2B) show the typical bands of a silica xerogel. The bands at 800 cm$^{-1}$ and 1080 cm$^{-1}$ correspond to Si—O symmetric and antisymmetric stretching vibrations from siloxane groups [22], whereas the shoulder at 1161 cm$^{-1}$ is associated to Si—O antisymmetric stretching vibrations in high molecular weight siloxane chains [23]. The presence of the –CH$_2$– and –CH$_3$ $\nu_s$ (2922 and 2852 cm$^{-1}$), $\nu_a$ bands (1400–1480 cm$^{-1}$) and $\delta$ H-C-H (1296 cm$^{-1}$) [23,24] indicates the presence of non-hydrolyzed ethoxy-groups and/or trapped ethanol in the pores, in accordance with the previous observations, which showed a weight loss lower than theoretical values. The band appearing at 970 cm$^{-1}$ is characteristic of the Si—O stretching vibration of Si-OH groups [25], formed after the hydrolysis, that did not undergo the condensation stage.

Thermogravimetric analysis confirmed that the C—H bands in the FTIR spectra (Fig. 2B) result from two contributions. Specifically, the TGA curve (Fig. 2C) shows two weight losses typically found on silica xerogels [26]. A first loss of 30% occurred at lower temperatures (90–360 °C), associated with the evaporation of reaction byproducts ($H_2O$, EtOH) trapped in the pores or physisorbed on the surface as evidenced by the signals at m/c = 18 ($H_2O$) and m/c = 31 (EtOH) by MS analysis of released products (Fig. S1A). The second weight loss, representing an 8%, occurred at 460–590 °C, and corresponds to the decomposition of non-hydrolyzed ethoxy groups from the precursor, manifested by the release of $CO_2$ (m/c = 44) and $H_2O$, tracked by MS analysis.

The progress of the hydrolysis reaction could be estimated by recalculating the weight losses with respect to the initial sol weight, considering simultaneously the values obtained by TGA and the 35% weight loss that occurred from the initial sol to the xerogel (calculations are detailed in supporting material). The recalculated data indicate a 39% $SiO_2$ content (practically the same as the theoretical value), a 5.2% loss associated to unreacted ethoxy groups and a combined 54.5% loss in hydrolysis byproducts ($H_2O$ and EtOH). The degree of hydrolysis, calculated (see eq. 1) as the relation between the experimental weight loss in hydrolysis byproducts and the theoretical weight loss that would result from a fully realized reaction (60%), was estimated at 91%, indicating that the majority of the silica precursor was hydrolyzed during the process.

$$Degree\ of\ hydrolysis = \frac{Experimental\ Loss\,(\%)}{Theoretical\ Loss\,(\%)} \times 100 \qquad (1)$$

The MAS-$^{29}$Si NMR experiments were employed to study the reticulation degree of the silica structure. The NMR spectra (Fig. 2D) showed the presence of four characteristic signals [27,28], corresponding to the various coordination degrees of the Si atom. The −111 and −102 ppm signals correspond to the Q$^4$ (Si(O–)$_4$ tetrahedra) and Q$^3$ (Si(OEt)(O–)$_3$) species respectively, which correspond respectively to the three-dimensional SiO$_2$ network and the in-plane Si–O units. The width of these bands may indicate a contribution of the signals associated with Q$^3$ (Si(OH)(O–)$_3$), as also suggested by the silanol band at 970 cm$^{-1}$ observed in the FTIR spectra (Fig. 2B). Linear arrangements were also evidenced by the signals at −96 ppm, attributed to Q$^2$ (Si(OEt)$_2$(O–)$_2$) species, and −87 ppm, corresponding to Q$^1$ species (Si(OEt)$_3$(O–)$_1$). The presence of these species is in line with the incomplete degree of hydrolysis calculated by TGA (93%) and the C–H bands observed in the FTIR spectra. From the integrated areas under the different signals of the $^{29}$Si NMR spectra (Fig. 2D), 54% (Q$^4$), 22% (Q$^3$), 14% (Q$^2$) and 10% (Q$^1$), it can be deduced that the majority of Si–O units (76%) are part of reticulated structures, confirming the formation of the three-dimensional network typical of silica xerogels.

### 3.2. Characterization of the reaction products with portlandite

Due to the difficulty of studying all the different interactions of the UCA-T product with a system as complex as a cement paste (i.e. presence of multiple phases, heterogeneous pore structure, varying water content in the pores…), a first approach to study the interaction to form C-S-H gel, was to perform a series of experiments with portlandite paste (Ca(OH)$_2$ + H$_2$O). This model system was chosen considering that portlandite is one of the hydration phases of cement, and can act as a Ca$^{2+}$ source with basic character, both of them conditions which favor pozzolanic-like reactions with the silica source [29].

The X-ray diffractograms of the portlandite paste (PP) (Fig. 3A) showed the presence of two crystalline phases: portlandite (Ca(OH)$_2$) and a lower proportion of calcite (CaCO$_3$). The presence of calcite indicates a partial carbonation, which might have been produced during the sample preparation and analysis. The PP + UCA-T diffractograms presented a similar profile (calcite and portlandite as the main signals), which indicated that the reaction products are not crystalline. The only difference was a slight broadening of the bands at 2θ = 29–30° (Fig. 3B), which may be attributed the formation of a poorly crystalline phase different from the amorphous silica xerogel produced by auto-condensation of the silica oligomer (Fig. 2A). The Rietveld analysis of this signal showed a fit to the Tobermorite 9 Å phase [30], as observed by the improvement of the fit in Fig. S2 ($\chi^2$ decreases from 1.70 to 1.38). As described by different models, tobermorite crystalline structure can be related to the short-term ordered domains of the C-S-H gel structure [31].

It is worth mentioning that the data obtained by XRD are insufficient to confirm the nature of the possible reaction products (i.e. SiO$_2$ and C-S-H gels), since they are predominantly amorphous and, therefore, they will produce weak and broad signals, which are masked by the peaks of the crystalline phases.

The elemental composition of the portlandite and portlandite + UCA-T pastes was analyzed by XRF (Table S2). PP was mainly composed of Ca phases (ca. 98%), with a minor amount of Mg, Si and Al impurities. The portlandite + UCA-T pastes, showed the presence of Si phases (ca. 10%, expressed as SiO$_2$). It is remarkable that the experimental % SiO$_2$ is lower than the nominal SiO$_2$ proportion (assuming that UCA-T contains 40% SiO$_2$) added to the paste (ca. 14.6%), which may indicate that a fraction of the silica oligomers did not react and were washed out during the solvent exchange with ethanol. The composition did not significantly vary from the paste cured for 7 and 21 days.

The reactivity of the UCA-T sol with Ca(OH)$_2$ to form C-S-H-like products was evidenced by the FTIR spectra (Fig. 4). The spectra of the Ca(OH)$_2$ paste showed a characteristic intense and narrow band at 3640 cm$^{-1}$, attributable to the $\nu_{as}$ O–H vibration in the Ca(OH)$_2$ structure, along with the characteristic CO$_3^{2-}$ bands at 870 ($\delta$ C–O) and 1450 cm$^{-1}$ ($\nu_{as}$ C–O), in line with the calcite content found by XRD measurements (Fig. 4A). The PP + UCA-T paste spectra showed some evident differences to the pure portlandite as a result of the reactions. First, the area ratio between the O–H portlandite band and the internal standard C≡N band (O-H/C ≡ N) decreased from ~2 to ~1 while the C=O/C ≡ N ratio (associated with carbonation) remained roughly the same, confirming that a fraction of the portlandite had been consumed during the reaction with UCA-T. No significant changes in portlandite or carbonate proportion were observed from 7 to 21 days. In addition to these changes, the characteristic $\nu_{as}$ H-O-H bands around 3200–3500 cm$^{-1}$ became evident in the portlandite + UCA-T paste spectra. The presence of this band can be explained by different contributions: (i) adsorbed H$_2$O due to a higher surface area, (ii) hydration H$_2$O molecules integrated between the Ca–O layers of the C-S-H structure and (3) the presence of Si-OH groups. The intensity of this band decreased from 7 to 21 days, which may result from two factors: (i) a reduction in the amount of adsorbed water and (ii) the transformation of silanols into silicates as the reaction with Ca(OH)$_2$ progresses (or into SiO$_2$ after auto-condensation).

The most noteworthy change observed in the spectra was the presence of a band group associated to silica, centered at 960–970 cm$^{-1}$. The position of these bands is characteristic of $\nu_{as}$ Si–O stretching vibrations ($\nu_{as}$) of the tetrahedral units forming part of linear chains (Q$^2$), present in the C-S-H structure [32]. It should be noted that the position of the Si–O bands differs from the FTIR spectra of the UCA-T xerogel (see Fig. 2B), where the bands at 1080 and 800 cm$^{-1}$ clearly indicated the formation of a three-dimensional SiO$_2$ network, thus confirming that the presence of Ca$^{2+}$ ions and a basic media alters the polymerization route of the UCA-T sol, yielding linear silica chains which are consistent with the structures of C-S-H gel [33]. An evolution of the structure from 7 to 21 days was evidenced by a shift of the band maximum from 957 to 969 cm$^{-1}$, which can be associated to the formation of a C-S-H structure with lower Ca/Si ratios [32]. This suggests that the hydrolyzed SiO$_4^-$ groups from the impregnation product are able to incorporate into the C-S-H silicate chain and decrease its Ca/Si ratio, which is consistent with the increase of the area ratio between the silica band and the internal standard (C≡N) band from 3.1 to 5.2. Another difference with respect to the UCA-T xerogel spectra was the total absence of the C–H bands in the presence of portlandite, which is a consequence of the strongly basic pH in the reaction mixture promoting a higher hydrolysis rate.

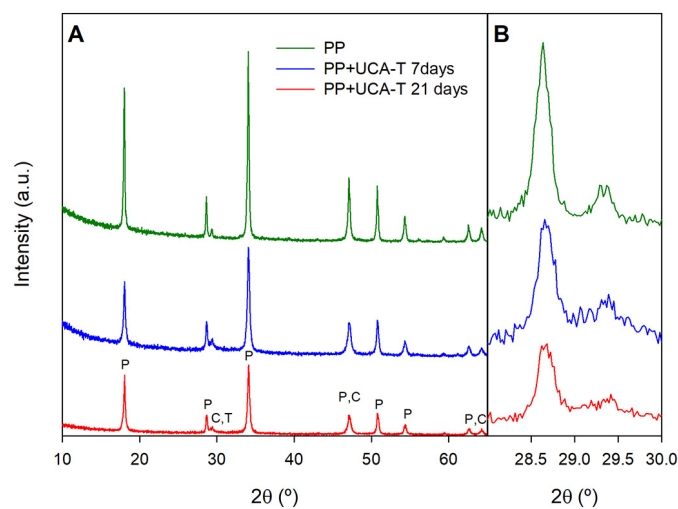

**Fig. 3.** (A) XRD diffractograms of the PP and PP + UCA-T pastes cured for 7 and 21 days. (B) Inset showing the region corresponding to the C-S-H signals. P: Portlandite (COD 9000013), C: Calcite (COD 7022027), T: Tobermorite (COD 9005447).
COD: Crystallography Open Database, http://cod.crystallography.net.

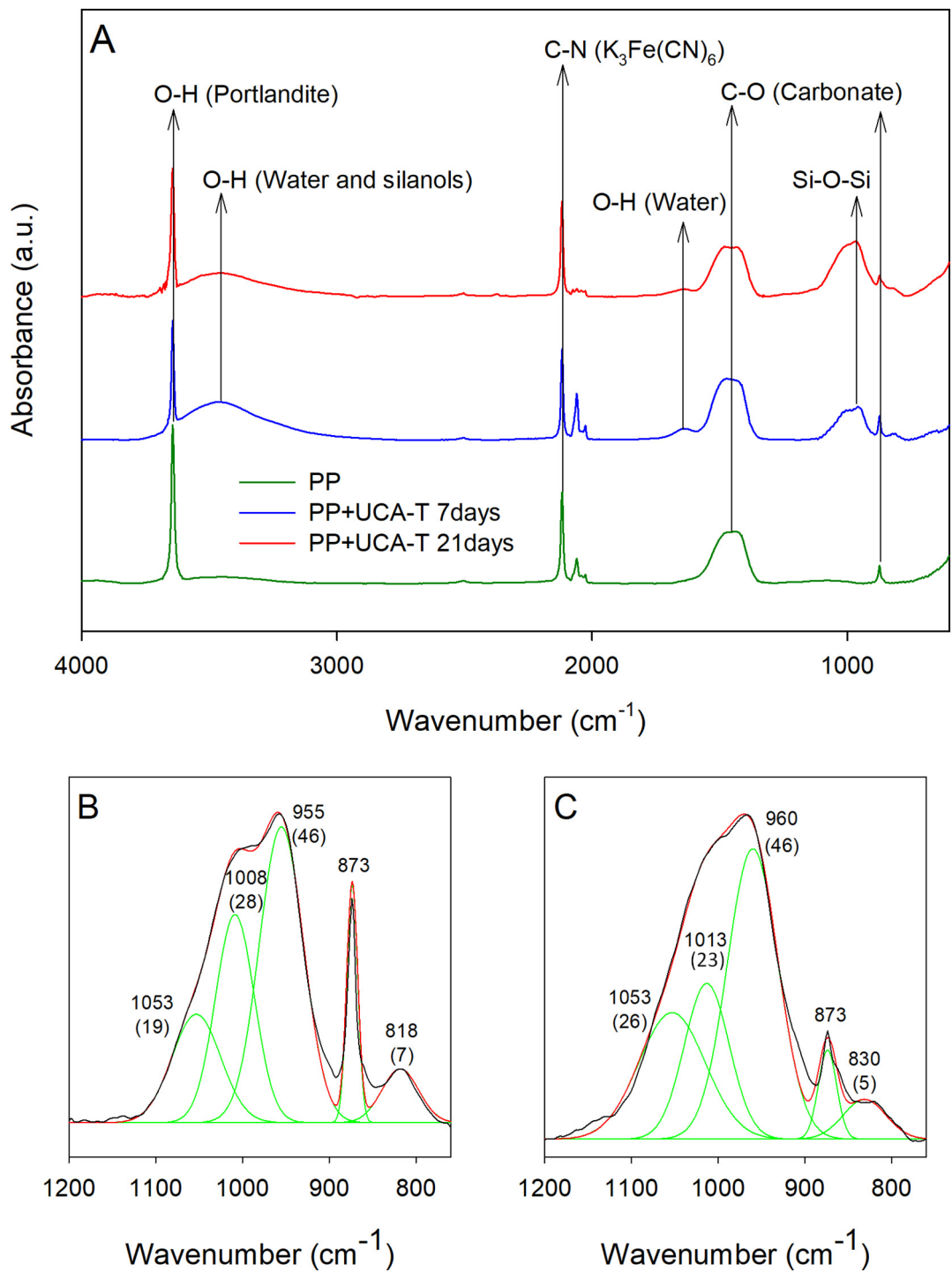

**Fig. 4.** (A) FTIR spectra of the portlandite paste (PP) cured for 7 days and PP + UCA-T cured for 7 and 21 days. (B, C) Curve fitting of the Si-O-Si band group showing the band positions and the percentage of contribution of Si–O bands of PP + UCA-T cured for (B) 7 days and (C) 21 days. *$K_3Fe[CN]_6$ was added as an internal standard.

The curve fitting analysis of the 760–1200 region (Fig. 4B, C) showed one carbonate band at 873 cm$^{-1}$ and four contributions of Si−O bands: (1) the Si−O stretching of Q1 tetrahedra at 820 cm$^{-1}$; (2) the asymmetric $\upsilon$(Si−O) $Q^2$ band centered around 960 cm$^{-1}$; (3) a band at 1013 cm$^{-1}$, which can be attributed to formation of silicates with a higher connectivity in Ca poor gels [34]; and (4) a band around 1053 cm$^{-1}$ corresponding to the reticulated $SiO_2$ networks ($Q^3$ and $Q^4$). The predominance of linear structures becomes evident by the higher contribution of the bands at 960 and 1013 cm$^{-1}$ respect to the 1053 cm$^{-1}$ signal. The curve fitting analysis further evidenced the evolution of the system from 7 to 21 days by a shift of the silicate bands group towards higher wavenumbers, which is indicative of a higher polymerization degree of the silica chains in C-S-H [35], and the lower contribution of the 820 cm$^{-1}$ ($Q^1$) signal. Furthermore, a decrease in the contribution of the 1013 cm$^{-1}$ indicated that part of the silicates have evolved towards structures with higher connectivity while the linear chains (960 cm$^{-1}$) were mostly retained.

An analysis of the phases present in the pastes was carried out by thermogravimetric analysis. In addition to the initial loss of weakly adsorbed water, the TGA curves (Fig. 5A) showed weight losses associated to three different processes: (1) 25–330 °C, evaporation of $H_2O$, adsorbed and hydration water in C-S-H, (2) 330–480 °C, $Ca(OH)_2 \rightarrow CaO + H_2O \uparrow$, (3) 520–660 °C, $CaCO_3 \rightarrow CaO + CO_2 \uparrow$. The assignment of these processes is corroborated by the evolution of the $H_2O$ (m/c = 18) and EtOH (m/c = 46) signals of evolved gases followed by MS analysis (Fig. S1B).

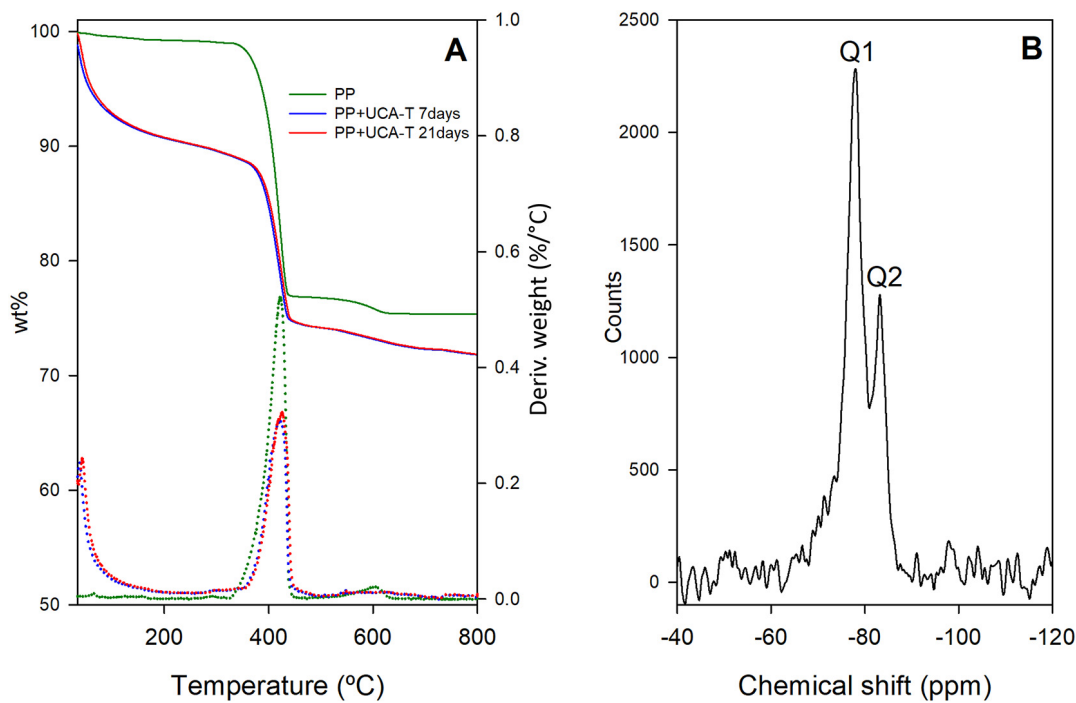

**Fig. 5.** (A) TGA and DTG curves of the portlandite paste with and without UCA-T cured for 7 and 21 days. (B) $^{29}$Si NMR of the portlandite + UCA-T paste cured for 21 days.

The detailed quantification results of the different components, calculated from the stoichiometry of the reactions, can be seen in Table S3. The portlandite paste contained a 90% $Ca(OH)_2$ and 3% $CaCO_3$. After the reaction with UCA-T, the pastes presented a higher content in $H_2O$, which can be associated to the hydration water in C-S-H and/or adsorbed $H_2O$ due to its high surface area. The proportion of $Ca(OH)_2$ significantly decreased in the portlandite + UCA-T pastes, down to values below 60%. These values are smaller than the nominal content of $Ca(OH)_2$ added to the initial mixture (81%, disregarding $H_2O$ and assuming total hydrolysis of the silica oligomer), thus confirming its reaction with the UCA-T product. The increase in %$CaCO_3$ was small, indicating that carbonation barely contributed to the consumption of $Ca(OH)_2$. Further evidence of the reaction of the silica oligomer with $Ca(OH)_2$ can be found in the proportion of the residual weight, which represented a 25–30% of the sample. The fact that these values were higher than the $SiO_2$ content measured by XRF (ca. 10.5%) can only be explained if we consider that those phases incorporate Si and Ca (e.g. C-S-H), in contrast with the practically pure $SiO_2$ content of the xerogel formed in absence of portlandite.

With regard to the evolution of the PP + UCA-T paste over time, variations in the phase proportions were barely observed. Specifically, the residue (silica phases) and the hydration water of C-S-H increased by < 0.2 while $Ca(OH)_2$ proportion decreased by < 0.1%, indicating a rapid equilibration of the $Ca(OH)_2$ reaction with the silica source. The small magnitude of the compositional changes seems to suggest that the evolution of the paste from 7 to 21 days is mostly the result of a restructuring of the calcium silicate phases.

The $^{29}$Si NMR spectra, measured for the PP + UCA-T paste cured for 21 days (Fig. 5B), confirmed the different structure of the silica units compared to the xerogel. Specifically, the portlandite + UCA-T paste exclusively presented peaks corresponding to $Q^2$ (−85 ppm) and $Q^1$ (−79 ppm) positions, indicating that the silica tetrahedral units were predominantly forming the linear chains characteristic of C-S-H, as also deduced by the position of the Si−O bands in the FTIR spectra (Fig. 4). The mean chain length (MCL) of the CSH gel in the PP + UCA-T, calculated according to eq. 2 [36], was 2.6 units, which is intermediate between the typical dimeric and pentameric units from jennite and tobermorite respectively [37]. This value is consistent with C-S-H gels with a Ca/Si ratio in the range 1.2–1.7 [38], specifically, similar MCL values were observed for CSH gels obtained from lime and silica using a mechanochemical process with Ca/Si ratios of 1.16 and 1.24 [36].

$$MCL = 2\frac{I(Q1) + I(Q2)}{I(Q1)} \qquad (2)$$

Another important difference with respect to the UCA-T xerogel lies in the shift of the $Q^2$ and $Q^1$ band positions. As discussed for the xerogel (Fig. 2D), the positions −96 ppm and −87 ppm correspond to Si $(OEt)_2(O-)_2$ and $Si(OEt)_3(O-)$ respectively [27], as a result of the incomplete hydrolysis of the silica oligomers. On the other hand, in the presence of portlandite, the signals appeared in positions corresponding to the Si atoms bonded to the CaO tetrahedra at chain-end (−78 ppm) and mid-chain (−86 ppm) positions. The position of these signals confirms the hydrolysis of the silica oligomer incorporated into the structure, as was suggested by the absence of C−H bands in the FTIR spectra.

In addition to the chemical and structural characterization of the formed C-S-H gel, the evolution of the morphological features of the reaction products was studied by SEM. The SEM study (Fig. 6) revealed a gradual evolution from the bare portlandite towards the growth of characteristic C-S-H structures over the $Ca(OH)_2$ surface. Without the addition of UCA-T (Fig. 6A, B), the portlandite paste presented a structure consisting on agglomerated crystals of varying size (0.25–3 μm) with a flat surface and the occasional presence of smaller rhombohedral crystals (50–100 nm) corresponding to calcite, as also detected by other techniques (XRD, TGA, FTIR). After 1 day curing in the presence of UCA-T (Fig. 6C, D), a porous globular-like structure, with 50–100 nm pores, over the Ca(OH)$_2$ grains became evident. This newly formed phase had a completely different structure than the compact packing of 30–70 nm particles of the UCA-T xerogel (see Fig. 1A, B), confirming the previous observations which suggested a change in the polymerization route of the silica oligomer. This structure observed here resembles that of unhydrated sol-gel derived calcium silicate cements observed in works from other authors [39].

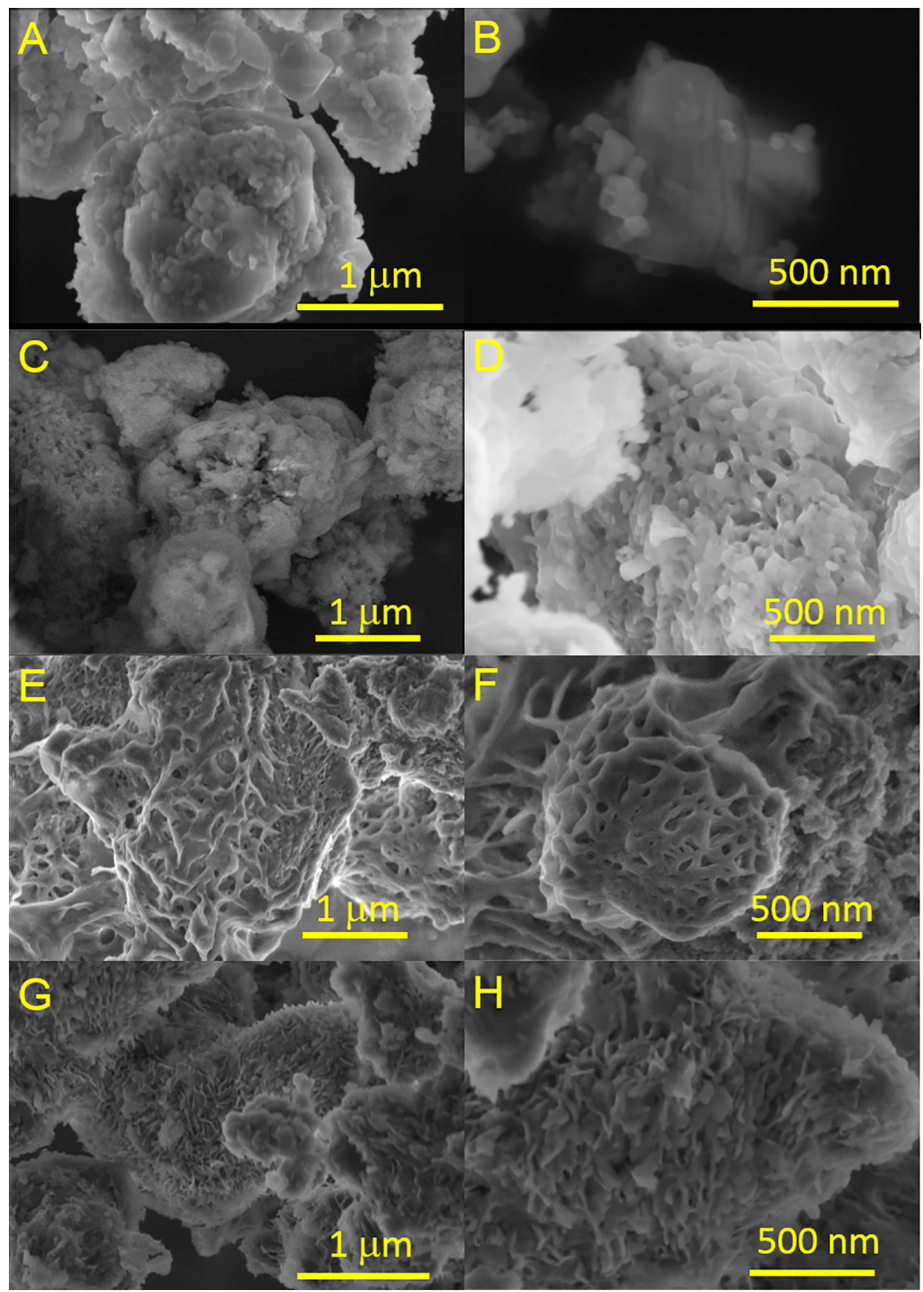

**Fig. 6.** SEM micrographs of (A, B) portlandite paste cured 7 days; portlandite paste + UCA-T cured (C, D) for 1 day; (E,F) for 7 days and (G,H) for 21 days.

After 7 days (Fig. 6E, F), the morphology of the structures evolved towards a sheet-like structure with larger (100–200 nm pores) and the coverage of the portlandite crystals was practically total, indicating the formation of calcium silicate progressed after 1 day. Albeit poorly defined, the morphological features of the reaction products at this stage showed some resemblance to the early hydration phases of alite (C$_3$S) [19,40], suggesting that from day 1 to day 7 the calcium silicate phases had started to undergo the hydration process, causing a re-arrangement of its structure into the characteristic sheet morphology of C-S-H gel. This explanation is supported by the previously discussed characterization techniques, which clearly showed a reaction product with C-S-H chemical/structural features at this stage.

The evolution of the reaction products towards more defined structures became clear after 21 days (Fig. 6E, F). More specifically, it was observed that the sheet-like structures grew and transformed into thin (< 20 nm) flake-like structures. These morphological features closely resemble some of the structures of C-S-H reported by other authors for C$_3$S pastes cured in limewater solution [19], synthetic C-S-H prepared via sol-gel routes [41] or pozzolanic reaction of portlandite with nano-silica [41]. The formation of these structures is consistent with the work by Zhang et.al [19], which reported that, when cured in submerged conditions, sheet-like structures were favored over the needle-like C-S-H due to the dilution of Ca$^{2+}$ ions in the surrounding liquid, leading to a reticular network structure. It should also be mentioned that the coverage of the Ca(OH)$_2$ crystals was practically the same as it was observed at 7 days suggesting that during the 7–21 days interval the main processes occurring were the hydration and a rearrangement of the structure, instead of the formation of new calcium silicate.

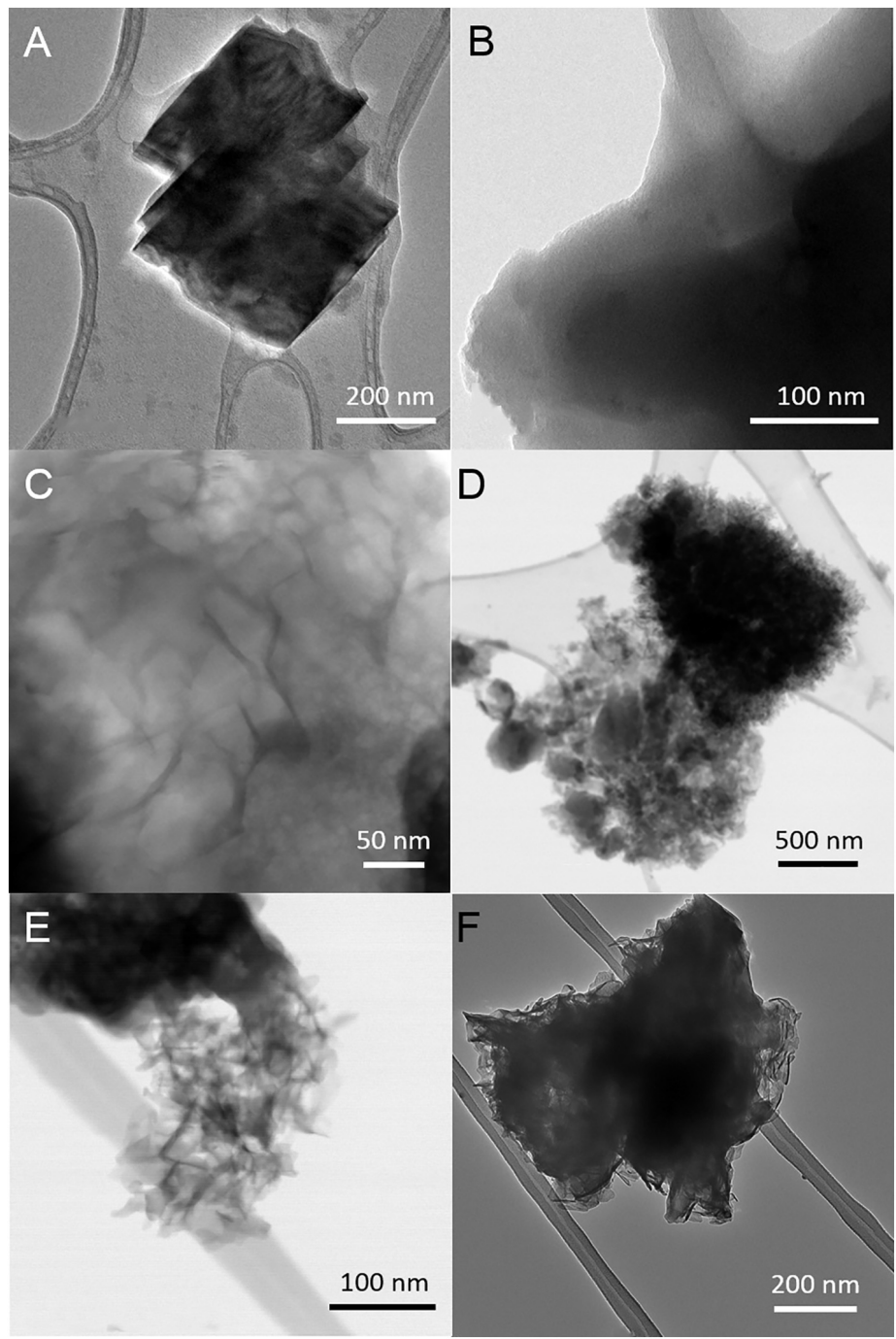

**Fig. 7.** TEM and STEM-BF micrographs of (A) portlandite paste cured 7 days and portlandite + UCA-T paste (B) cured for 1 day, (C, D) cured for 7 days, (E,F) cured for 21 days.

In a similar vein to the SEM images, the TEM micrographs (Fig. 7) showed, in further detail, the formation of the characteristic sheet-like structures of C-S-H after 7 days, which became more defined after 21 days. The PP + UCA-T paste at 1 day showed the formation of a structure without defining features over the Ca(OH)$_2$ crystals. Analysis by EDX of these amorphous structures showed that the Ca/Si ratio greatly varied depending on the measured region, from 8.5 to 0.1 values, indicating a heterogeneous phase composition (i.e. coexistence of calcium silicate, Ca(OH)$_2$, calcite, SiO$_2$ xerogel remnants…). After curing for 7 days, the measured Ca/Si ratios of the C-S-H structures varied between 2.3 and 3.0, which are slightly higher than the typical ratios reported for calcium-rich C-S-H gels [36]. It should be noted that, due to the sensitivity of the structure to the electron beam, the analysis was performed with a large spot size and the Ca signal may be overestimated by the proximity of the portlandite crystals.

After curing for 21 days, the Ca/Si ratio evolved towards values in the 1.0–1.4 range, which fit into the characteristic proportions found in C-S-H. This decrease in the Ca/Si ratio can be explained by the increase in the degree of polymerization of the C-S-H silica chains, in line with the FTIR measurements, which showed a shift of the silicate bands towards higher wavenumbers (Fig. 4) and a decreased contribution of the Q$^1$ Si band (~810–820 cm$^{-1}$ in the FTIR spectra).

In light of the results obtained from the different techniques, we propose that the reaction of the product with the Ca(OH)$_2$ takes places through the following stages: (i) the strongly basic media produces the hydrolysis and cleavage of the silica oligomers during the first stages of the process, yielding reactive silanol groups, which start to react with the calcium released in the vicinity of the Ca(OH)$_2$ crystals to form calcium silicate hydrate. (ii) This reaction continues through the first seven days, coexisting with the growth of the C-S-H gel structures. (iii) From 7 to 21 days, the condensation continues and the silica chains in C-S-H undergo further polymerization, causing a re-arrangement of the structure.

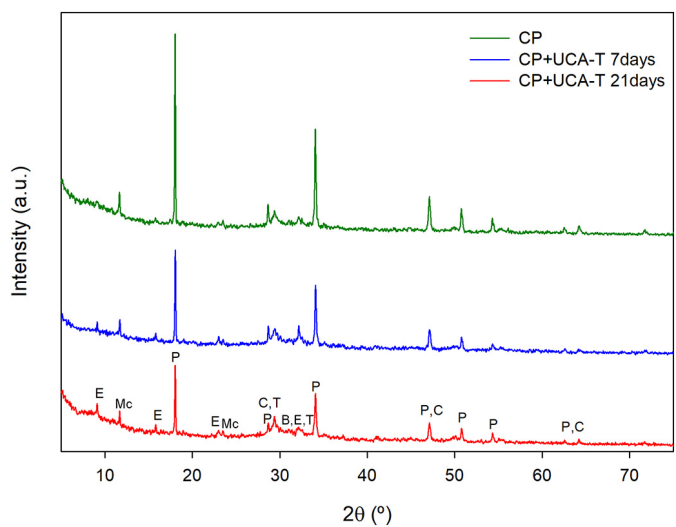

**Fig. 8.** XRD diffractograms of the CP and CP + UCA-T pastes cured for 7 and 21 days and quantification of crystalline phase by Rietvald method. P: Portlandite (COD 9000013), C: Calcite (COD 7022027), T: Tobermorite (COD 9005447), E: Ettringite (COD 9015084), Mc: C4AcH11 monocarboaluminate (COD 2007668), B: Belite (COD 1535815).
COD: Crystallography Open Database, http://cod.crystallography.net.

### 3.3. Characterization of the reaction products with hydrated cement paste

Having demonstrated the ability of the UCA-T product to react with portlandite, the following stage of the study was focused on the evaluation of the interactions with hydrated cement paste (CP).

Crystallographic analysis by XRD (Fig. 8) identified different characteristic phases in the cement paste. The most abundant crystalline phase was portlandite, followed by ettringite and monocaboaluminate (AFm), resulting from the hydration of aluminum phases. Additionally, the analysis showed the minor presence of remnant belite (C$_2$S), and carbonation products (calcite). A weaker broad signal appeared at 2θ = 29–30°, which can be associated to the short-ordered domains of C-S-H, resulting from the hydration of C$_2$S and C$_3$S. The XRD patterns of the CP + UCA-T displayed the same phases, although with differences in the relative peak intensities, confirming that the reaction products were predominantly amorphous phases. In general, it was observed that the intensity of the portlandite signal decreased with respect to the other phases while the calcite peaks barely increased, evidencing that the portlandite is consumed by the reaction with the product instead of carbonation processes. No discernable differences in the crystalline composition were found between the CP + UCA-T cured for 7 and 21 days.

Elemental analysis by XRF of the CP showed a typical composition of cement (see Table S4), mainly composed of CaO (69%) and SiO$_2$ (18%), and a smaller proportion of Al and Fe (7%) oxides. In the case of the CP + UCA-T pastes, the SiO$_2$ content increased to ~30–31% (differences between the pastes at 7 and 21 days were not significant), indicating the integration of the product in the system. This value practically matches the nominal proportion (ca. 31.9%), suggesting that the hydrolysis rate is faster in the presence of cement paste than in the Ca(OH)$_2$ paste, where evidence of unreacted oligomers was observed.

The nature of the reaction products was further studied by analysis of the FTIR spectra (Fig. 9). The CP presented the characteristic $\upsilon_{O-H}$ (3300–3500 cm$^{-1}$) and $\delta_{O-H}$ (~1600 cm$^{-1}$) bands of H$_2$O, corresponding to the water molecules of the hydrated phases (C-S-H, ettringite, AFm…). The narrow O−H band at 3640 cm$^{-1}$ was indicative of the presence of portlandite. The broad signal at ~1420 cm$^{-1}$ can be attributed to $\nu_{as}$ C−O of calcite and monocarboaluminates. Finally, a broad band group appeared in the 800–1300 cm$^{-1}$ range, comprising a shoulder at 1111 cm$^{-1}$ attributed to $\nu_{as}$ S−O antisymmetric stretching vibrations of SO$_4^{2-}$ groups from ettringite, a broad signal at 960 cm$^{-1}$ corresponding to Si−O antisymmetric stretching vibrations of SiO$_4^{4-}$ groups from CSH gel and a shoulder at 875 cm$^{-1}$ asigned to δ C−O (CO$_3^{2-}$) and a small shoulder at 810 cm$^{-1}$ corresponding to $\nu_s$Si-O Q1 in CSH gel. The CP + UCA-T spectra showed the same bands, but with some differences in their relative intensity. More specifically, the band at 3640 cm$^{-1}$ decreased in intensity (as also observed in the portlandite + UCA-T pastes) due to the consumption of portlandite in the reaction with the silica oligomers. The increased intensity of the bands corresponding to hydration H$_2$O, along with the band group associated with the silicates evidenced the formation of new C-S-H. The intensity of the latter bands slightly increased from day 7 to day 21, indicating the progress of the reaction. It is also worth mentioning that the absence of C−H bands confirmed the full hydrolysis of the silica oligomer.

The curve fitting analysis of the 750–1300 cm$^{-1}$ band group (Fig. 9B, C, D) revealed the contribution of five different signals. The signal at 1100 cm$^{-1}$ can be attributed to the $\upsilon_3$(S−O) from ettringite. The ~870 cm$^{-1}$ band corresponds to $\upsilon_2$(C−O) from carbonates. The silica bands showed three contributions: the characteristic C-S-H bands [32] at ~970 cm$^{-1}$ and 820 cm$^{-1}$, respectively associated with Q2(Si−O) and Q1(Si−O) from the silicate chain, and a band with a low contribution at ~1040 cm$^{-1}$ corresponding to Q3 and Q4 Si−O units. The presence of the latter is likely a result of the carbonation process of C-S-H.

After the reaction with UCA-T, the contribution of the 1104 cm$^{-1}$ decreased overall, indicating the formation of Si−O structures and/or the reaction of the UCA-T product with the ettringite. The quantification of the peak areas revealed an increase of the contribution of the Q2 Si−O band (~970 cm$^{-1}$) and a slight decrease of the Q1 Si−O band (~820 cm$^{-1}$) respect to the CP, which may be indicative of a lengthening of the silica chains in C-S-H. This fact was further evidenced by the shift of the bands towards higher wavelengths, in line with the observations made for the PP + UCA-T pastes (Fig. 4). In a similar vein, the band associated with three-dimensional silica structures (~1040 cm$^{-1}$) shifted towards higher wavelength and showed a slightly higher contribution. Nevertheless, the low value of the latter suggests that the main reaction product is C-S-H gel instead of the SiO$_2$ xerogel.

Regarding the evolution of the reaction from 7 to 21 days, it can be observed that the Si−O (Q2) and Si−O (Q1) bands slightly increased, indicating the formation of new C-S-H gel during this interval. The position of these bands remained roughly the same, which can be interpreted as an evidence that the silica chains did not undergo further polymerization, unlike in the 0–7 days interval. On the other hand, the fact that the 1040 cm$^{-1}$ (Si−O networks) and 876 cm$^{-1}$ (CO$_3^{2-}$) bands did not increase suggests that the SiO$_2$ xerogel formation and carbonation processes may be considered insignificant under the experimental conditions.

A more detailed characterization of the phases present in the pastes was performed by thermogravimetric analysis. The TGA curves of the CP and CP + UCA-T (Fig. 10) showed the same three regions: (1) 25–360 °C, H$_2$O loss (free water, C-S-H, ettringite, AFm…) (2) 360–460 °C, Ca(OH)$_2$ → CaO + H$_2$O ↑, (3) 520–670 °C, CaCO$_3$ → CaO + CO$_2$ ↑. The DTG revealed two peaks attributed water loss: 26–110 °C and 110–200 °C. Detailed results of the quantification of the different components, calculated through the stoichiometry of the reactions, are available in Table S5.

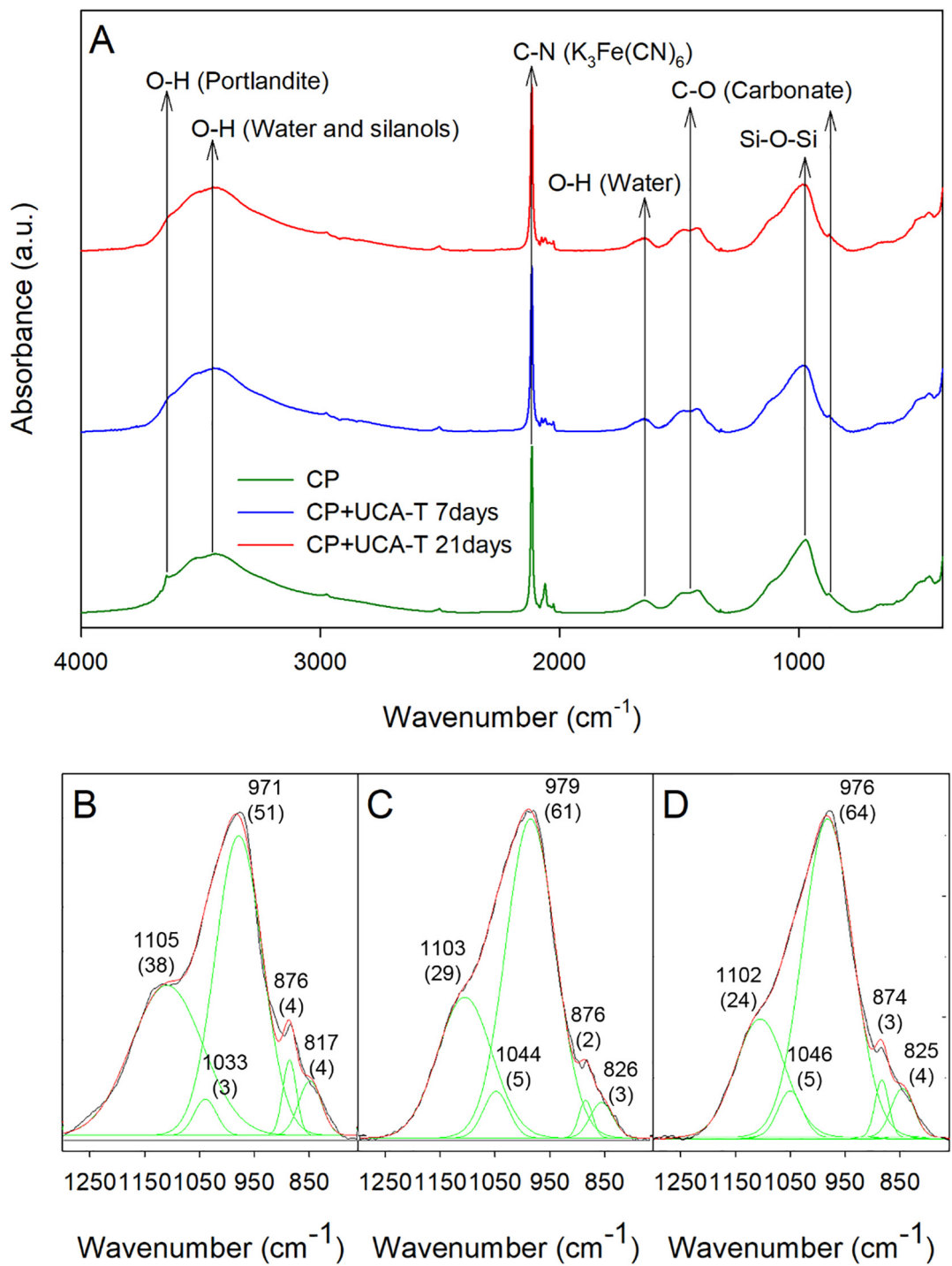

**Fig. 9.** (A) FTIR spectra of the CP and CP + UCA-T after 7 and 21 days. (B, D) Curve fitting of the Si-O-Si band group showing their position and percentage of contribution, (B) CP, (C) CP + UCA-T after 7 days and (D) CP + UCA-T after 21 days. *$K_3Fe[CN]_6$ was added as an internal standard.

In agreement with the FTIR spectra, the TG analysis evidenced a decrease in the proportion of $Ca(OH)_2$ from 24.2% in CP to 15.7% and 14.8% after 7 and 21 days mixed with UCA-T, respectively. These values are lower than the nominal proportion (c.a. 17.9%), thus confirming that the portlandite had been consumed during the process. The low magnitude of the increase in %$CaCO_3$ (< 2%) suggests that the main process consuming the portlandite was its reaction with UCA-T instead of the carbonation process. In a similar vein to the results from the $Ca(OH)_2$ pastes, the amount of water from hydrated phases increased from 14.1% to 15.9%, and there was an increase in the residual weight from 56.2% to ~61.5%, suggesting the formation of similar reaction products (i.e. C-S-H). The partial carbonation from 7 to 21 days, observed by FTIR, was also confirmed by a 1–2% increase in the $CaCO_3$ proportion.

Fig. 11 shows the $^{29}Si$ NMR spectra used for the structural characterization of CP and CP + UCA-T after curing for 21 days. The cement paste (CP) spectra presented the characteristic signals [36,42] of the $Q^1$ chain end (−79 ppm) and $Q^2$ mid-chain (−85 ppm) silica tetrahedra constituting the linear chains in C-S-H, aside from the $Q^0$ signals (around −71 ppm) corresponding to the silica units present in the anhydrous phases (alite and belite). The presence of a shoulder around

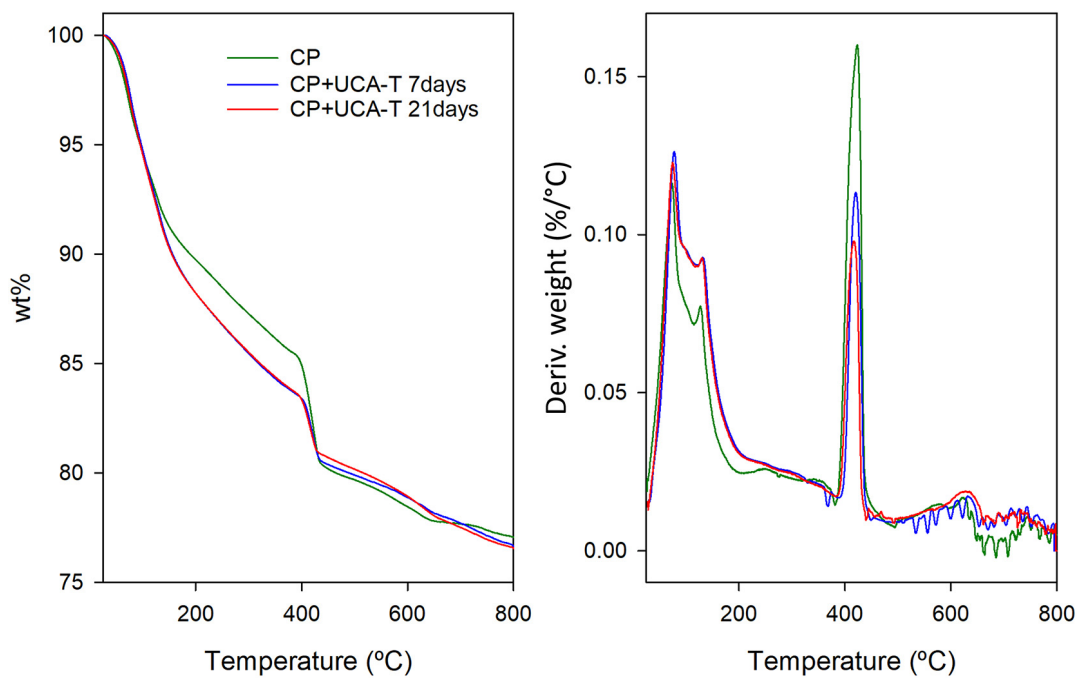

**Fig. 10.** TGA and DTG curves of CP and CP + UCA-T cured for 7 and 21 days.

−81 ppm can be attributed to mid-chain Si units with adjacent aluminum-containing ($Q^2(1Al)$) tetrahedra [43].

In the case of the CP + UCA-T sample, the spectra showed similar bands, with a minor contribution of the $Q^3$ signals, indicating that the product predominantly polymerized in the form of linear chains, instead of the typical reticulated structure of silica xerogels [10,44]. This observation can be the consequence of two different phenomena: (1) the silica oligomers react with the portlandite present in the cement paste to form C-S-H, as demonstrated by the experiments with the portlandite paste and the FTIR and TGA results with CP. (2) The product gets incorporated into the silica chains of the existing C-S-H structure, as evidenced by an increased intensity of the $Q^2$ signal respect to $Q^1$.

A more detailed analysis of the structural changes was carried out through curve fitting analysis of the $^{29}$Si NMR signals (Fig. 11C, D and Table 1). In the case of the plain CP, the $Q^0$ band resulted from the contribution of two signals at −71.3 and − 73.60 ppm corresponding to belite (β-C2S) and alite (Si(5)) respectively. The $Q^2$ signal from mid-chain units in C-S-H (−84.6 ppm) had a lower intensity compared to the $Q^1$ band (−78.9 ppm). The length of the silica chains (MCL) was estimated as 4.6 Si−O units, by using the equation proposed by Richardson [43], which takes into account the presence of aluminum-containing units (Eq. (3)).

$$MCL = 2\frac{\left(Q1 + Q2 + Q2(l) + \frac{3}{2}Q2(1Al)\right)}{Q1} \quad (3)$$

Significant changes were observed in the CP after reaction with UCA-T that indicated the integration of the silica oligomer in the structure. On the one hand, the contribution of the $Q^0$ signals decreased, indicating a progress of the hydration process. On the other hand, the contribution of the $Q^2$ and $Q^1$ signals is inverted ($Q^2 > Q^1$) compared to the paste without UCA-T. The global contribution of the $Q^2(1Al)$ units decreased, although the $Q^2(1Al)/Q^1$ area ratio was roughly the same as in the paste without UCA-T. As a result of the contribution of $Q^2$ units, the MCL increased from 4.6 to 6.0 units. This observation evidences that the silica tetrahedral units resulting from the UCA-T hydrolysis may be incorporated into the C-S-H gel [36,42]. The contribution of the $Q^3$ signals was barely significant (4%), indicating that the xerogel formation was not favored.

The morphological features of the reaction products were observed by Scanning Electron microscopy. Micrographs of the plain CP (available in supporting information Fig. S3) showed the characteristic morphologies of the hydrated phases detected by other techniques: (1) the flake-like structures of Type II C-S-H. (2) Ettringite needles. (3) Hexagonal crystals of portlandite. After the reaction with UCA-T, the paste showed similar features in general, and the formation of new C-S-H structures was difficult to detect considering that this phase is already one of the main components of the CP. Nevertheless two new structures were observed, with globular and fibrous morphologies respectively. The globular structure can be attributed to the formation of smaller proportions of silica xerogel, as also observed in the FTIR spectra (Fig. 8). The fibrous structures may be related with the newly formed C-S-H by reaction with the portlandite. The formation of these morphologies instead of the flake-like structures is probably a result of the different curing conditions after mixing with UCA-T, thus promoting a different growth mechanism. This observation is in agreement with other works, where it has been found that curing in a high %RH atmosphere leads to the formation of fiber-like (Type I) C-S-H structures in cement pastes [19] due to the "whisker growth" mechanism, where water evaporation from the surface causes the accumulation of ions in the solution which favors the overgrowth of C-S-H.

## 4. Conclusions

We have demonstrated the ability of a silica oligomer based-impregnation treatment, developed in our laboratory, to produce C-S-H gel in contact with $Ca^{+2}$ ions existing in porlandite, one of the components of cementitious materials. A thorough characterization of the products resulting from the interaction between silica oligomer and cementitious matrix components demonstrated that the polymerization of the silica precursor occurs through a different pathway in their presence yielding reaction products with the structural, chemical and morphological properties of C-SH gel, instead of the reticulated silica xerogel that is produced by auto-polymerization of silica oligomer chains.

Specifically, when polymerization occurred in the presence of portlandite and water, the silica precursor reacted with the solubilized

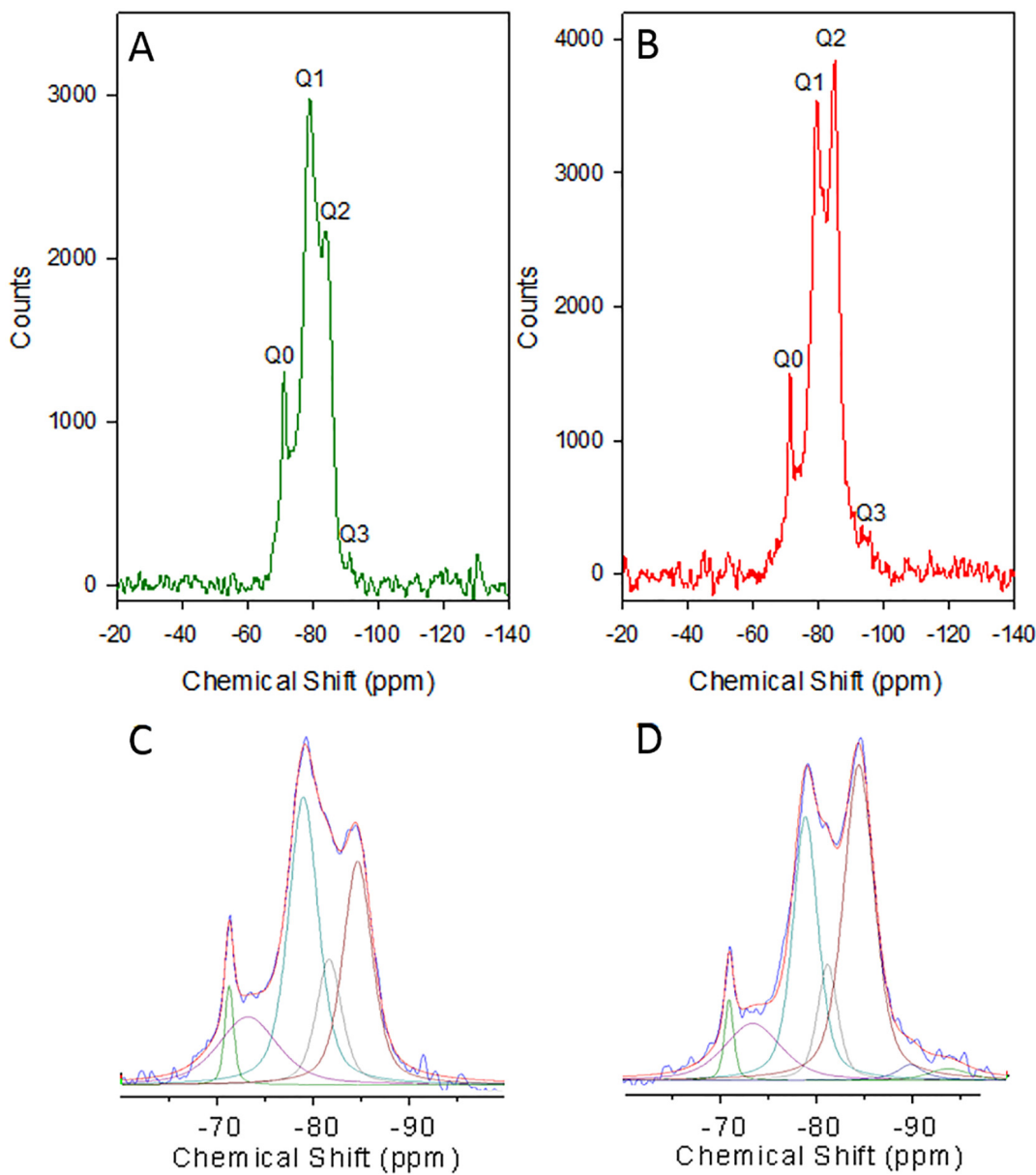

**Fig. 11.** MAS $^{29}$Si NMR spectra of (A) CP and (B) CP + UCA-T after 21 days. (C,D) Curve fitting analysis of the respective $^{29}$Si NMR signals.

$Ca^{2+}$ ions, as evidenced by a gradual decrease in the amount of portlandite, giving rise to an amorphous phase (according to XRD) with silica units polymerized into linear chains (according to $^{29}$Si NMR and FTIR) and a MCL of 3.5, in agreement with the typical structural units of C-S-H. At early stages (1 day), the newly formed phase showed a non-distinctive shape, which evolved in time (at 21 days) to a foil-like morphology due to hydration processes. The integration of the silica units into the CaO structure and the formation of the hydrated phase was further confirmed by the evolution of the Ca/Si ratio from > 8.0 to ~1.0 values and by the shift of the FTIR bands through the reaction time.

The experiments with the product and hydrated cement paste mixtures showed the formation of reaction products with similar characteristics to the ones formed after reaction with portlandite paste by two simultaneous phenomena: (i) the silica oligomer reacts with the free portlandite present in the cement paste (evidenced by TGA, XRD and FTIR). (ii) The silica units are incorporated into the existing C-S-H structure, resulting in an increased average chain length and the enhancement of $Q^2$ Si−O units in bridging positions (as observed by $^{29}$Si NMR). The presence of reticulated chains and structures with xerogel morphology was marginal, indicating that the reaction with the paste is favored respect to the auto-condensation. Further studies should be carried out in order to determine the possible changes in reactivity with cement pastes affected by chemical degradation processes that decrease the portlandite content and alter the C-S-H structure (e.g. carbonation).

**Table 1**
Position and contribution (in Area %) of the bands from the $^{29}$Si NMR spectra of CP and CP + UCA-T pastes.

| Sample | | Clinker | | Reaction product | | | | | MCL |
|---|---|---|---|---|---|---|---|---|---|
| | | $Q^0$ | | $Q^1$ | $Q^2$(1Al) | $Q^2$ | $Q^3$ | | |
| CEMENT PASTE | δ (ppm) | −71.23 | −73.19 | −78.97 | −81.64 | −84.64 | – | – | 4.65 |
| | % | 4 | 17 | 37 | 14 | 28 | – | – | |
| CEMENT PASTE + UCA-T | δ (ppm) | −71.15 | −73.60 | −79.10 | −81.40 | −84.66 | −90.02 | −94.00 | 6.00 |
| | % | 3 | 14 | 28 | 10 | 41 | 2 | 2 | |

In light of the obtained results, the similarity of the reaction products with the cementitious matrix indicates a high chemical compatibility with concrete and mortars.

### CRediT authorship contribution statement

**Rafael Zarzuela:** Investigation, Formal analysis, Writing - original draft, Writing - review & editing. **Manuel Luna:** Investigation, Formal analysis, Writing - original draft, Writing - review & editing. **Luis M. Carrascosa:** Investigation. **María P. Yeste:** Investigation, Writing - review & editing. **Inés Garcia-Lodeiro:** Methodology, Resources, Formal analysis, Writing - review & editing. **M. Teresa Blanco-Varela:** Methodology, Resources, Formal analysis, Writing - review & editing. **Miguel A. Cauqui:** Investigation, Writing - review & editing. **José M. Rodríguez-Izquierdo:** Writing - review & editing, Supervision, Project administration. **María J. Mosquera:** Writing - review & editing, Supervision, Project administration.

### Declaration of competing interest

The authors declare that they have no known competing financial interests or personal relationships that could have appeared to influence the work reported in this paper.

### Acknowledgements

This project has received funding from the European Union's Horizon 2020 - Research and Innovation Framework Programme under grant agreement No 760858; and from Spanish Government/FEDER-EU under research project MAT2017-84228-R.

### Appendix A. Supplementary data

Supplementary data to this article can be found online at https://doi.org/10.1016/j.cemconres.2020.106008.

### References

[1] F. Sandrolini, E. Franzoni, H. Varum, R. Nakonieczny, Materiales y tecnologías en la Arquitectura Modernista: Casos de Estudio de decoración de fachadas en Italia, Portugal y Polonia persiguiendo una restauración racional, Inf. La Construcción. 63 (2011) 5–11, https://doi.org/10.3989/ic.10.053.

[2] Q.T. Phung, N. Maes, D. Jacques, J. Perko, G. De Schutter, G. Ye, Modelling the evolution of microstructure and transport properties of cement pastes under conditions of accelerated leaching, Constr. Build. Mater. 115 (2016) 179–192, https://doi.org/10.1016/j.conbuildmat.2016.04.049.

[3] V.H. Nguyen, B. Nedjar, J.M. Torrenti, Chemo-mechanical coupling behaviour of leached concrete, Nucl. Eng. Des. 237 (2007) 2090–2097, https://doi.org/10.1016/j.nucengdes.2007.02.012.

[4] A. Custance-baker, S. Macdonald, Conserving Concrete Heritage Experts Meeting, June 9–11 2014 The Getty Conservation Institute, Los Angeles, California, 2015.

[5] X. Pan, Z. Shi, C. Shi, T.-C. Ling, N. Li, A review on concrete surface treatment part I: types and mechanisms, Constr. Build. Mater. 132 (2017) 578–590, https://doi.org/10.1016/J.CONBUILDMAT.2016.12.025.

[6] P. Hou, X. Cheng, J. Qian, S.P. Shah, Effects and mechanisms of surface treatment of hardened cement-based materials with colloidal nanoSiO$_2$ and its precursor, Constr. Build. Mater. 53 (2014) 66–73, https://doi.org/10.1016/j.conbuildmat.2013.11.062.

[7] G. Wheeler, Alkoxysilanes and the Consolidation of Stone, The Getty Conservation Institute, Los Angeles, California, 2005, https://doi.org/10.1007/s13398-014-0173-7.2.

[8] J.F. Illescas, M.J. Mosquera, Producing surfactant-synthesized nanomaterials in situ on a building substrate, without volatile organic compounds, ACS Appl. Mater. Interfaces 4 (2012) 4259–4269, https://doi.org/10.1021/am300964q.

[9] J.F. Illescas, M.J. Mosquera, Surfactant-synthesized PDMS/silica nanomaterials improve robustness and stain resistance of carbonate stone, J. Phys. Chem. C 115 (2011) 14624–14634, https://doi.org/10.1021/jp203524p.

[10] D.S. Facio, M. Luna, M.J. Mosquera, Facile preparation of mesoporous silica monoliths by an inverse micelle mechanism, Microporous Mesoporous Mater. 247 (2017) 166–176, https://doi.org/10.1016/j.micromeso.2017.03.041.

[11] R. Zarzuela, M. Luna, L.A.M. Carrascosa, M.J. Mosquera, Preserving cultural heritage stone: innovative consolidant, superhydrophobic, self-cleaning, and biocidal products, Adv. Mater. Conserv. Stone, Springer International Publishing, Cham, 2018, pp. 259–275, , https://doi.org/10.1007/978-3-319-72260-3_12.

[12] Mosquera, M.J.; Illescas, J.F.; Facio, D.S. International Patent. No. WO2013/121058, February 14, 2014.

[13] F. Elhaddad, L. Carrascosa, M. Mosquera, Long-term effectiveness, under a mountain environment, of a novel conservation nanomaterial applied on limestone from a Roman archaeological site, Materials (Basel) 11 (2018) 694, https://doi.org/10.3390/ma11050694.

[14] B. Pigino, A. Leemann, E. Franzoni, P. Lura, Ethyl silicate for surface treatment of concrete – part II: characteristics and performance, Cem. Concr. Compos. 34 (2012) 313–321, https://doi.org/10.1016/J.CEMCONCOMP.2011.11.021.

[15] E. Franzoni, B. Pigino, C. Pistolesi, Ethyl silicate for surface protection of concrete: performance in comparison with other inorganic surface treatments, Cem. Concr. Compos. 44 (2013) 69–76, https://doi.org/10.1016/j.cemconcomp.2013.05.008.

[16] S. Goñi, F. Puertas, M.S. Hernández, M. Palacios, A. Guerrero, J.S. Dolado, B. Zanga, F. Baroni, Quantitative study of hydration of C3S and C2S by thermal analysis, J. Therm. Anal. Calorim. 102 (2010) 965–973, https://doi.org/10.1007/s10973-010-0816-7.

[17] F. Sandrolini, E. Franzoni, B. Pigino, Ethyl silicate for surface treatment of concrete - part I: pozzolanic effect of ethyl silicate, Cem. Concr. Compos. 34 (2012) 306–312, https://doi.org/10.1016/j.cemconcomp.2011.12.003.

[18] A.M. Barberena-Fernández, P.M. Carmona-Quiroga, M.T. Blanco-Varela, Interaction of TEOS with cementitious materials: chemical and physical effects, Cem. Concr. Compos. 55 (2015) 145–152, https://doi.org/10.1016/j.cemconcomp.2014.09.010.

[19] Z. Zhang, G.W. Scherer, A. Bauer, Morphology of cementitious material during early hydration, Cem. Concr. Res. 107 (2018) 85–100, https://doi.org/10.1016/j.cemconres.2018.02.004.

[20] T. Gerber, B. Himmel, C. Hübert, WAXS and SAXS investigation of structure formation of gels from sodium water glass, J. Non-Cryst. Solids 175 (1994) 160–168, https://doi.org/10.1016/0022-3093(94)90008-6.

[21] S.L.B. Lana, A.B. Seddon, X-ray diffraction studies of sol-gel derived ORMOSILs based on combinations of tetramethoxysilane and trimethoxysilane, J. Sol-Gel Sci. Technol. 13 (1998) 461–466, https://doi.org/10.1023/A:1008685614559.

[22] L. Téllez, J. Rubio, E. Morales, J.L. Oteo, Synthesis of inorganic organic hybrid materials from TEOS, TBT and PDMS.pdf, J. Mater. Sci. (2003) 1773–1780.

[23] Z. Demjen, B. Pukanszky, E. Foldes, J. Nagy, Interaction of silane coupling agents with CaCO$_3$, J. Colloid Interface Sci. 190 (1997) 427–436, https://doi.org/10.1006/jcis.1997.4894.

[24] P. Innocenzi, Infrared spectroscopy of sol-gel derived silica-based films: a spectra-microstructure overview, J. Non-Cryst. Solids 316 (2003) 309–319, https://doi.org/10.1016/S0022-3093(02)01637-X.

[25] M.J. Mosquera, D.M. De Los Santos, A. Montes, L. Valdez-Castro, New nanomaterials for consolidating stone, Langmuir 24 (2008) 2772–2778, https://doi.org/10.1021/la703652y.

[26] A. Fina, D. Tabuani, F. Carniato, A. Frache, E. Boccaleri, G. Camino, Polyhedral oligomeric silsesquioxanes (POSS) thermal degradation, Thermochim. Acta 440 (2006) 36–42, https://doi.org/10.1016/J.TCA.2005.10.006.

[27] F. Devreux, J.P. Boilot, F. Chaput, A. Lecomte, Sol-gel condensation of rapidly hydrolyzed silicon alkoxides: a joint Si29 NMR and small-angle x-ray scattering study, Phys. Rev. A 41 (1990) 6901–6909, https://doi.org/10.1103/PhysRevA.41.6901.

[28] Z. Sassi, J.C. Bureau, A. Bakkali, Structural characterization of the organic/inorganic networks in the hybrid material (TMOS–TMSM–MMA), Vib. Spectrosc. 28 (2002) 251–262, https://doi.org/10.1016/S0924-2031(01)00158-8.

[29] A. Moropoulou, A. Cakmak, K.C. Labropoulos, R. Van Grieken, K. Torfs, Accelerated microstructural evolution of a calcium-silicate-hydrate (C-S-H) phase in pozzolanic pastes using fine siliceous sources: comparison with historic pozzolanic mortars, Cem. Concr. Res. 34 (2004) 1–6, https://doi.org/10.1016/S0008-8846(03)00187-X.

[30] S. Merlino, E. Bonaccorsi, T. Armbruster, Tobermorites; their real structure and order-disorder (OD) character, Am. Mineral. 84 (1999) 1613–1621, https://doi.org/10.2138/am-1999-1015.

[31] S. Papatzani, K. Paine, J. Calabria-Holley, A comprehensive review of the models on the nanostructure of calcium silicate hydrates, Constr. Build. Mater. 74 (2015) 219–234, https://doi.org/10.1016/j.conbuildmat.2014.10.029.

[32] P. Yu, R.J. Kirkpatrick, B. Poe, P.F. McMillan, X. Cong, Structure of calcium silicate hydrate (C-S-H): near-, mid-, and far-infrared spectroscopy, J. Am. Ceram. Soc. 82 (2004) 742–748, https://doi.org/10.1111/j.1151-2916.1999.tb01826.x.

[33] R.J. Kirkpatrick, J.L. Yarger, P.F. Mcmillan, P. Yu, Raman spectroscopy of C-S-H, Tobermorite, and Jennite (1997) 7355, https://doi.org/10.1016/S1065-7355(97)00001-1.

[34] J. Björnström, A. Martinelli, A. Matic, L. Börjesson, I. Panas, Accelerating effects of colloidal nano-silica for beneficial calcium-silicate-hydrate formation in cement, Chem. Phys. Lett. 392 (2004) 242–248, https://doi.org/10.1016/j.cplett.2004.05.071.

[35] M. Monasterio, J.J. Gaitero, E. Erkizia, A.M. Guerrero Bustos, L.A. Miccio,

J.S. Dolado, S. Cerveny, Effect of addition of silica- and amine functionalized silica-nanoparticles on the microstructure of calcium silicate hydrate (C-S-H) gel, J. Colloid Interface Sci. 450 (2015) 109–118, https://doi.org/10.1016/j.jcis.2015.02.066.
[36] E. Tajuelo Rodriguez, I.G. Richardson, L. Black, E. Boehm-Courjault, A. Nonat, J. Skibsted, Composition, silicate anion structure and morphology of calcium silicate hydrates (C-S-H) synthesised by silica-lime reaction and by controlled hydration of tricalcium silicate ($C_3S$), Adv. Appl. Ceram. 114 (2015) 362–371, https://doi.org/10.1179/1743676115Y.0000000038.
[37] E. Tajuelo, K. Garbev, D. Merz, L. Black, I.G. Richardson, Cement and concrete research thermal stability of C-S-H phases and applicability of Richardson and Groves' and Richardson C-(A)-S-H (I) models to synthetic C-S-H, Cem. Concr. Res. 93 (2017) 45–56, https://doi.org/10.1016/j.cemconres.2016.12.005.
[38] I.G. Richardson, Model structures for C-(A)-S-H(I), Acta Crystallogr. Sect. B Struct. Sci. Cryst. Eng. Mater. 70 (2014) 903–923, https://doi.org/10.1107/S2052520614021982.
[39] B.S. Lee, H.P. Lin, J.C.C. Chan, W.C. Wang, P.H. Hung, Y.H. Tsai, Y.L. Lee, A novel sol-gel-derived calcium silicate cement with short setting time for application in endodontic repair of perforations, Int. J. Nanomedicine 13 (2018) 261–271, https://doi.org/10.2147/IJN.S150198.
[40] R. Ylmén, U. Jäglid, B.M. Steenari, I. Panas, Early hydration and setting of Portland cement monitored by IR, SEM and Vicat techniques, Cem. Concr. Res. 39 (2009) 433–439, https://doi.org/10.1016/j.cemconres.2009.01.017.
[41] Q. Lin, X. Lan, Y. Li, Y. Ni, C. Lu, Y. Chen, Z. Xu, Preparation and characterization of novel alkali-activated nano silica cements for biomedical application, J. Biomed. Mater. Res. - Part B Appl. Biomater. 95 B (2010) 347–356, https://doi.org/10.1002/jbm.b.31722.
[42] I. García-Lodeiro, A. Fernández-Jiménez, I. Sobrados, J. Sanz, A. Palomo, C-S-H gels: interpretation of 29Si MAS-NMR spectra, J. Am. Ceram. Soc. 95 (2012) 1440–1446, https://doi.org/10.1111/j.1551-2916.2012.05091.x.
[43] I.G. Richardson, Nature of C-S-H in hardened cements, Cem. Concr. Res. 29 (1999) 1131–1147, https://doi.org/10.1016/S0008-8846(99)00168-4.
[44] A. Depla, D. Lesthaeghe, T.S. Van Erp, A. Aerts, K. Houthoofd, F. Fan, C. Li, V. Van Speybroeck, M. Waroquier, C.E.A. Kirschhock, J.A. Martens, $^{29}$Si NMR and UV-raman investigation of initial oligomerization reaction pathways in acid-catalyzed silica Sol-Gel chemistry, J. Phys. Chem. C 115 (2011) 3562–3571, https://doi.org/10.1021/jp109901v.